\documentclass[journal]{IEEEtran}
\usepackage{amsmath,amsfonts}
\usepackage{array}
\usepackage[caption=false,font=normalsize,labelfont=sf,textfont=sf]{subfig}
\usepackage{textcomp}
\usepackage{bm}
\usepackage{stfloats}
\usepackage{algorithm}
\usepackage{algpseudocode}
\usepackage{booktabs}
\usepackage{placeins}
\usepackage{url}
\usepackage{amsmath,amssymb}
\usepackage{verbatim}
\usepackage{graphicx}
\def\BibTeX{{\rm B\kern-.05em{\sc i\kern-.025em b}\kern-.08em
    T\kern-.1667em\lower.7ex\hbox{E}\kern-.125emX}}
\usepackage{balance}
\usepackage{xcolor}
\begin{document}
\title{\textcolor{black}{Structured PARAFAC Channel Estimation\\ for Flexible Intelligent Metasurface Receivers}}
\author{Xi Han\textcolor{black}{\textsuperscript{1,2}}, Junqiao Chen\textcolor{black}{\textsuperscript{1}}, Yuansheng Zhang\textcolor{black}{\textsuperscript{2,*}}, André L. F. de Almeida\textcolor{black}{\textsuperscript{3,*}},\\ and George C. Alexandropoulos\textcolor{black}{\textsuperscript{4}}
\thanks{\textcolor{black}{\textsuperscript{1}X. Han and J. Chen are with the Beijing Key Laboratory on Integration and Analysis of Large-scale Stream Data, School of Information Science and Technology, North China University of Technology, Beijing 100144, China (e-mail:jinlucky333@126.com; cjq1251726586@gmail.com).} \textcolor{black}{\textsuperscript{2}X. Han and Yuansheng Zhang are with the State Key Laboratory of Intelligent Optimized Manufacturing in Mining \& Metallurgy Process (e-mail: zhangyuansheng@126.com).} \textcolor{black}{\textsuperscript{3}André L. F. de Almeida is with the Laboratory of Applied Signal Processing and the Wireless Telecom Research Group, Department of Teleinformatics Engineering, Federal University of Ceará, Fortaleza, Brazil (e-mail: andre@gtel.ufc.br).} \textcolor{black}{\textsuperscript{4}G. C. Alexandropoulos is with the Department of Informatics and Telecommunications, National and Kapodistrian University of Athens, 16122 Athens, Greece (e-mail: alexandg@di.uoa.gr).} \textcolor{black}{\textsuperscript{*}Corresponding authors: André L. F. de Almeida and Yuansheng Zhang.}}
\thanks{\textcolor{black}{This work was supported in part by the National Key Research and Development Program of China under Grant 2024YFE0200103, the China Scholarship Council (Grant No. 202508110189), the NSFC under Grant 62001008, and the Beijing Municipal Natural Science Foundation under Grants 4212002 and L192034. This work is partially supported by the National Institute of Science and Technology (INCT-Signals) sponsored by Brazil's National Council for Scientific and Technological Development (CNPq) (406517/2022-3), and FUNCAP (INCT-25255-82587.32.41/64). The research of André L. F. de Almeida is partially supported by CNPq (303356/2025-1).}}
}

\maketitle

\begin{abstract}
\textcolor{black}{Flexible intelligent metasurfaces (FIMs) create geometric diversity through aperture morphing, but their configuration-dependent responses complicate uplink channel estimation. This letter considers multiple users sharing a common angular response and develops a structured PARAFAC model across calibrated FIM configurations. Its known identity factor and rank-one pilot-weighted channel factor enable deterministic, noniterative algebraic factor recovery through per-element rank-one approximations, reference-state normalization, global rank-one projection, and a least-squares morphing-factor update. The recovered spatial and morphing factors are then jointly matched to the calibrated manifold to estimate the common angles, user gains, and configuration-dependent channels. Simulations demonstrate accurate recovery and show that out-of-plane morphing resolves the azimuth-sign ambiguity of an undeformed planar aperture; an angular-spread experiment quantifies the mismatch when the common-response assumption is relaxed.}
\end{abstract}


\renewcommand\baselinestretch{.85}

\section{Introduction}
\label{sec:introduction}

\IEEEPARstart{F}{lexible} intelligent metasurfaces (FIMs) introduce controllable geometric degrees of freedom by adjusting the positions of their constituent elements, thereby reshaping the effective aperture and modifying the spatial response \cite{An2025FIMMISO,An2025FIMMIMO}. This geometric reconfigurability provides additional propagation and spatial-diversity opportunities, but it also complicates channel acquisition because the array response varies with the adopted morphing configuration. Accurate channel estimation must therefore account for the coupling between the underlying propagation parameters and the prescribed FIM geometry.

Recent studies have investigated channel estimation for FIM and related geometry-reconfigurable architectures. Sparse angular recovery with FIM-shape optimization was considered in \cite{Jiang2026EnhancedFIM}, while model-based interpolation, kernel methods, sparse parametric recovery, and neural-operator learning were studied in \cite{Xiao2026FIMCE}. For movable-antenna MIMO systems, tensor decomposition was employed to exploit measurements collected over antenna movements \cite{Zhang2024MATensor}, and an alternating-least-squares PARAFAC formulation was introduced for point-to-point MIMO systems equipped with morphing FIM arrays \cite{Romano2026FIMPARAFAC}. \textcolor{black}{Tensor-based receiver designs have been investigated in UAV-aided MIMO systems, including information monitoring receivers \cite{Han2022UAVMonitoring}.} Tensor-based channel estimation has also been widely explored in systems incorporating reconfigurable intelligent surfaces (RISs) \cite{Araujo2021,ardah2021trice}, including PARAFAC models for RIS-assisted multiple-input multiple-output (MIMO) systems \cite{multiuseRIS,Cao2025TensorRIS} and low-rank tensor recovery for RIS-assisted millimeter-wave orthogonal frequency division multiplexing (OFDM) \cite{Zheng2022LowRankTensor}. \textcolor{black}{For multi-RIS uplink MIMO systems, a PARATUCK2-based framework was developed for joint channel and symbol estimation, followed by subspace-based geometric parameter extraction \cite{Han2026MultiRIS}.} \textcolor{black}{Structured tensor factorizations have also been investigated for beyond-diagonal RIS architectures \cite{Almeida2025BDRISTensor}.} These studies establish the value of tensor modeling, while motivating estimators that explicitly exploit the structure of multiuser uplinks observed across calibrated morphing states.

To address this need, we study uplink channel estimation for multiple users observed over prescribed, calibrated FIM morphing configurations. We focus on the common-angular-response regime, in which users share a dominant arrival direction as seen by the FIM; this model also approximates compact user clusters unresolved by the effective aperture. Relative to the ALS-based point-to-point FIM-PARAFAC formulation in \cite{Romano2026FIMPARAFAC}, we target the multiuser uplink and exploit the structure it induces: the known identity factor decouples the mode-1 unfolding into per-element rank-one matrix approximations, while the pilot-weighted channel factor obeys a global rank-one constraint. We use these properties for reference-state normalization, global rank-one projection, and a single least-squares morphing-factor refinement, avoiding alternating PARAFAC updates. We then jointly match the recovered spatial and morphing factors to the calibrated FIM manifold to estimate the common angles, user gains, and configuration-dependent channels. Simulations demonstrate accurate factor recovery and show how morphing resolves the azimuth-sign ambiguity inherent to the undeformed planar aperture. A nonzero-angular-spread experiment additionally quantifies the model mismatch as this assumption is relaxed.

\section{System Model and FIM-PARAFAC Formulation}
\label{sec:system_model}

\subsection{Multi-Morphing Uplink Model}
\begin{figure}[t]
	\centering
	\includegraphics[
	width=\columnwidth,
	trim=20 25 20 20,
	clip
	]{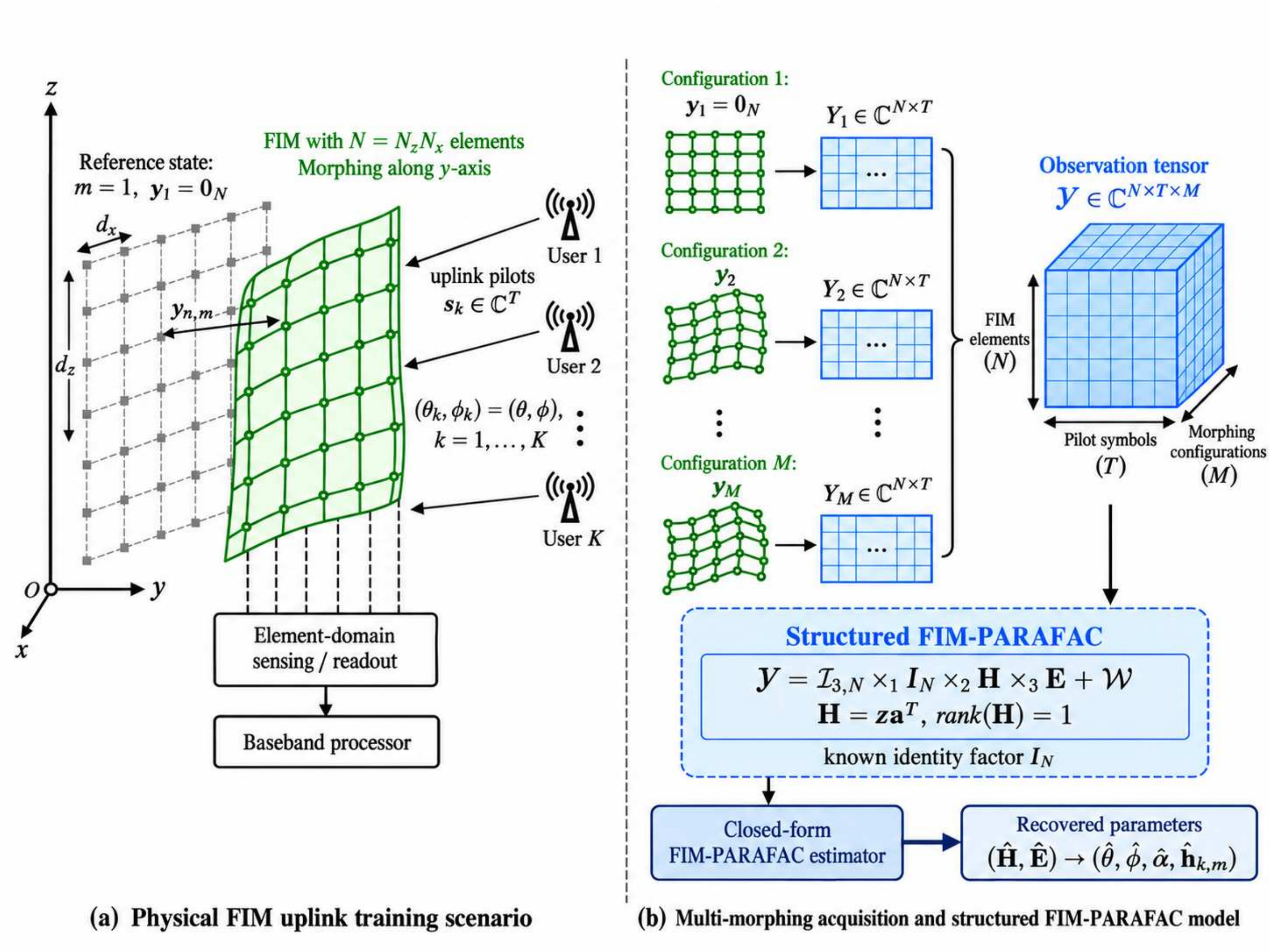}
	\caption{FIM-assisted uplink training and structured
		PARAFAC formulation.}
	\label{fig:system_model}
\end{figure}
As illustrated in Fig.~1, we consider a narrowband block-fading uplink with $K$ single-antenna users and an $N$-element FIM receiver. In the undeformed reference state, the $N=N_xN_z$ elements form a planar array in the $xz$-plane, indexed by $n=(p-1)N_z+q$, with $x_p=(p-1)d_x$ and $z_q=(q-1)d_z$. During training, the FIM cycles through $M$ prescribed, calibrated morphing configurations. Element positions remain fixed along $x$ and $z$ and vary only along $y$; thus, the position of element $n$ in configuration $m$ is \textcolor{black}{$\mathbf r_{n,m}=[x_p,\;y_{n,m},\;z_q]^T$},
\textcolor{black}{where $y_{n,m}$ is the element's out-of-plane displacement.}
The displacement profile of the $m$th configuration is collected as \textcolor{black}{$\mathbf y_m=[y_{1,m},\ldots,y_{N,m}]^T\in\mathbb R^N$}.

Consider a far-field plane wave with elevation angle $\theta$ and
azimuth angle $\phi$, with unit direction vector
\textcolor{black}{$\mathbf u(\theta,\phi)=[\sin\theta\cos\phi,\;\sin\theta\sin\phi,\;\cos\theta]^T$}.
Let $\lambda$ denote the carrier wavelength and
$\kappa=2\pi/\lambda$ the wavenumber. The response of the $n$th FIM
element under the $m$th morphing configuration is
\textcolor{black}{$[\mathbf a_m(\theta,\phi)]_n=e^{j\kappa\mathbf r_{n,m}^T\mathbf u(\theta,\phi)}$}.
Let the undeformed UPA steering vector be
\textcolor{black}{$\mathbf a_{\mathrm{UPA}}(\theta,\phi)=\mathbf a_x(\theta,\phi)\otimes\mathbf a_z(\theta)$},
and the morphing-induced phase vector be
\textcolor{black}{$\mathbf e_m(\theta,\phi)=e^{j\kappa\sin\theta\sin\phi\,\mathbf y_m}$},
where the exponential is applied elementwise. The complete FIM
steering vector is therefore
\textcolor{black}{$\mathbf a_m(\theta,\phi)=\mathbf a_{\mathrm{UPA}}(\theta,\phi)\odot\mathbf e_m(\theta,\phi)$}.
The first morphing configuration is selected as the undeformed
reference state, i.e.,
\begin{equation}
\mathbf{y}_{1}=\mathbf{0}_{N},
\qquad \mathbf{e}_{1}(\theta,\phi)=\mathbf{1}_{N}.
\label{eq:reference_state}
\end{equation}

During uplink training, user $k$ transmits a known pilot sequence
$\mathbf{s}_{k}\in\mathbb{C}^{T}$, and all pilot sequences are
collected in \textcolor{black}{$\mathbf S=[\mathbf s_1,\ldots,\mathbf s_K]^T\in\mathbb C^{K\times T}$}.
Let $\alpha_k\in\mathbb{C}$ denote the complex channel gain of user
$k$, and let $(\theta_k,\phi_k)$ denote its angular parameters.
Under the single-dominant-path model, the channel between user $k$
and the FIM under configuration $m$ is $\mathbf{h}_{k,m}
= \alpha_k\mathbf{a}_{m}(\theta_k,\phi_k)$.
The received pilot matrix under the $m$th morphing configuration is
therefore
\begin{equation}
\color{black}
\begin{aligned}
\mathbf{Y}_{m}
&=\sum_{k=1}^{K}\mathbf{h}_{k,m}\mathbf{s}_{k}^{T}
+\mathbf{W}_{m},\quad m=1,\ldots,M,
\end{aligned}
\label{eq:received_matrix}
\end{equation}
where
$\mathbf{W}_{m}\in\mathbb{C}^{N\times T}$ denotes additive noise.

In this work, we focus on the angularly co-located-user configuration,
for which \textcolor{black}{$\theta_k=\theta$ and $\phi_k=\phi$ for
$k=1,\ldots,K$}.
This condition applies exactly to users sharing the same dominant
arrival direction and also serves as a nominal model for a compact
user cluster that the effective FIM aperture cannot resolve.
Let \textcolor{black}{$\mathbf a\triangleq\mathbf a_{\mathrm{UPA}}(\theta,\phi)$},
\textcolor{black}{$\mathbf e_m\triangleq\mathbf e_m(\theta,\phi)$}, and
\textcolor{black}{$\boldsymbol\alpha=[\alpha_1,\ldots,\alpha_K]^T$}.
Define the pilot-weighted channel gain vector as
\textcolor{black}{$\mathbf z\triangleq\mathbf S^T\boldsymbol\alpha$}.
The signal model in \eqref{eq:received_matrix} is equivalent to
\begin{equation}
\color{black}
\mathbf{Y}_{m}=\operatorname{diag}\{\mathbf{e}_{m}\}\mathbf{a}\mathbf{z}^{T}+\mathbf{W}_{m}.
\label{eq:colocated_received}
\end{equation}

\subsection{Structured FIM-PARAFAC Representation}

The received matrices
$\{\mathbf{Y}_{m}\}_{m=1}^{M}$ are stacked along the morphing
configuration dimension to construct the third-order observation
tensor \textcolor{black}{$\mathcal Y\in\mathbb C^{N\times T\times M}$},
with \textcolor{black}{$\mathcal Y(:,:,m)=\mathbf Y_m$}.
Its three modes correspond to the FIM elements, pilot symbols, and
morphing configurations, respectively.

Define
\begin{equation}
\color{black}
\mathbf{H}\triangleq\mathbf{z}\mathbf{a}^{T}\in\mathbb{C}^{T\times N},\quad
\operatorname{rank}(\mathbf{H})=1,
\label{eq:H_rank_one}
\end{equation}
and collect the morphing responses in
\textcolor{black}{$\mathbf E\triangleq[\mathbf e_1,\ldots,\mathbf e_M]^T\in\mathbb C^{M\times N}$}.
Since the first configuration is undeformed, the reference condition
in \eqref{eq:reference_state} gives
\textcolor{black}{$\mathbf E(1,:)=\mathbf 1_N^T$}.

\textcolor{black}{Let $\mathcal{I}_{3,N}\in\mathbb{R}^{N\times N\times N}$ denote the third-order identity tensor. Substituting \eqref{eq:colocated_received} into the stacked observations gives the structured PARAFAC model}
\begin{equation}
\color{black}
\begin{aligned}
\mathcal{Y}
&=\mathcal{I}_{3,N}\times_{1}\mathbf{I}_{N}\times_{2}\mathbf{H}\times_{3}\mathbf{E}+\mathcal{W}.
\end{aligned}
\label{eq:structured_parafac}
\end{equation}
\textcolor{black}{where $\mathcal{W}$ is the stacked noise tensor. The factors are therefore $\mathbf{I}_{N}$, $\mathbf{H}$, and $\mathbf{E}$, with a known identity factor and the global rank-one constraint on $\mathbf{H}$ in \eqref{eq:H_rank_one}. Under the adopted unfolding convention, its mode-1 unfolding is}
\textcolor{black}{$[\mathcal Y]_{(1)}=(\mathbf E\diamond\mathbf H)^T+[\mathcal W]_{(1)}$}.
\textcolor{black}{Here, $\diamond$ denotes the Khatri--Rao product. Transposing this relation and defining}
\textcolor{black}{$\mathbf N_1\triangleq[\mathcal W]_{(1)}^T$}
\textcolor{black}{gives}
\begin{equation}
\color{black}
\mathbf{Q}\triangleq[\mathcal{Y}]_{(1)}^{T}=\mathbf{E}\diamond\mathbf{H}+\mathbf{N}_{1}.
\label{eq:mode1_model}
\end{equation}
The known identity factor and the columnwise structure of
\eqref{eq:mode1_model} enable the deterministic closed-form
factorization developed in the next section.

\section{Proposed Structured FIM-PARAFAC Estimator}
\label{sec:closed_form}

\subsection{Columnwise Rank-One Factorization}

Starting from the mode-1 representation in
\eqref{eq:mode1_model}, let
\begin{equation}
\mathbf{Q}
=
[\mathbf{q}_{1},\ldots,\mathbf{q}_{N}]
=
\mathbf{E}\diamond\mathbf{H}
+
\mathbf{N}_{1}
\in\mathbb{C}^{MT\times N}.
\label{eq:Q_definition}
\end{equation}
Denote by $\mathbf{h}_{n}$ and
$\boldsymbol{\epsilon}_{n}$ the $n$th columns of
$\mathbf{H}$ and $\mathbf{E}$, respectively. Since the
Khatri--Rao product operates columnwise, the $n$th column of
\eqref{eq:Q_definition} satisfies
\begin{equation}
\mathbf{q}_{n}
=
\boldsymbol{\epsilon}_{n}
\otimes
\mathbf{h}_{n}
+
\mathbf{n}_{n}.
\label{eq:column_model}
\end{equation}
Using
$\operatorname{vec}(\mathbf{h}_{n}
\boldsymbol{\epsilon}_{n}^{T})
=
\boldsymbol{\epsilon}_{n}\otimes\mathbf{h}_{n}$,
$\mathbf{q}_{n}$ can be reshaped into
\begin{equation}
\mathbf{Q}_{n}
\triangleq
\operatorname{unvec}_{T\times M}(\mathbf{q}_{n})
=
\mathbf{h}_{n}
\boldsymbol{\epsilon}_{n}^{T}
+
\mathbf{N}_{n},
\qquad
n=1,\ldots,N,
\label{eq:rankone_subproblem}
\end{equation}
where $\mathbf{N}_{n}$ denotes the corresponding reshaped
noise matrix. \textcolor{black}{Hence, the structured tensor
factorization is reduced to $N$ separate, algebraically decoupled
rank-one matrix approximation problems. This decoupling is algebraic
and does not, by itself, imply statistical independence of the noise
terms.}

Let
\begin{equation}
\mathbf{Q}_{n}
=
\mathbf{U}_{n}
\boldsymbol{\Sigma}_{n}
\mathbf{V}_{n}^{H}
\end{equation}
be the singular value decomposition of $\mathbf{Q}_{n}$, and
let $(\sigma_{n,1},\mathbf{u}_{n,1},\mathbf{v}_{n,1})$
denote its dominant singular triplet. A compatible pair of
unnormalized factor estimates is
\begin{equation}
\widetilde{\mathbf{h}}_{n}
=
\sqrt{\sigma_{n,1}}\mathbf{u}_{n,1},
\qquad
\widetilde{\boldsymbol{\epsilon}}_{n}
=
\sqrt{\sigma_{n,1}}\mathbf{v}_{n,1}^{*}.
\label{eq:svd_factor}
\end{equation}
The complex conjugation in
\eqref{eq:svd_factor} follows from the transpose in $\mathbf{h}_{n}\boldsymbol{\epsilon}_{n}^{T}$.
\textcolor{black}{The SVD factors in \eqref{eq:svd_factor} give the unweighted Frobenius-norm best rank-one approximation of each reshaped observation, $\widetilde{\mathbf h}_{n}\widetilde{\boldsymbol{\epsilon}}_{n}^{T}=\sigma_{n,1}\mathbf u_{n,1}\mathbf v_{n,1}^{H}$, with squared residual $\sum_{i=2}^{\min\{T,M\}}\sigma_{n,i}^{2}$. The minimizer is unique when its dominant singular value is simple, although its factor pair retains a reciprocal scalar ambiguity. Such an unweighted least-squares fit corresponds to a maximum-likelihood estimate only under an i.i.d. white Gaussian noise model.}
The two factors in \eqref{eq:svd_factor} are subject to a
reciprocal scaling ambiguity. This ambiguity is resolved by the
undeformed reference configuration in
\eqref{eq:reference_state}, which implies
$[\boldsymbol{\epsilon}_{n}]_{1}=1$. Defining
\begin{equation}
\beta_{n}
\triangleq
[\widetilde{\boldsymbol{\epsilon}}_{n}]_{1},
\end{equation}
the normalized estimates are obtained as
\begin{equation}
\widehat{\mathbf{h}}_{n}
=
\beta_{n}
\widetilde{\mathbf{h}}_{n},
\qquad
\widehat{\boldsymbol{\epsilon}}_{n}
=
\frac{
	\widetilde{\boldsymbol{\epsilon}}_{n}
}{\beta_{n}}.
\label{eq:column_normalization}
\end{equation}
Under the exact model, $\beta_n\neq0$ because the first entry
of each morphing-factor column equals one.
\textcolor{black}{With noise, $\beta_n\neq0$ remains a necessary
nondegeneracy condition, and a small $|\beta_n|$ makes this
reference-state normalization ill-conditioned. 
}

Collecting the normalized column estimates yields
\begin{equation}
\mathbf{H}_{\mathrm{CF}}
=
[\widehat{\mathbf{h}}_{1},
\ldots,
\widehat{\mathbf{h}}_{N}],
\qquad
\mathbf{E}_{\mathrm{CF}}
=
[\widehat{\boldsymbol{\epsilon}}_{1},
\ldots,
\widehat{\boldsymbol{\epsilon}}_{N}].
\label{eq:preliminary_factors}
\end{equation}

\subsection{Global Rank-One Projection and Factor Refinement}

\textcolor{black}{Since the matrices $\mathbf{Q}_{n}$ are approximated separately, their assembled estimate $\mathbf{H}_{\mathrm{CF}}$ will generally not satisfy the global physical constraint $\operatorname{rank}(\mathbf{H})=1$ in noise. We therefore enforce the common angular-response structure by}
\begin{equation}
\mathbf{H}_{\mathrm{r1}}
=
\mathcal{P}_{1}
\left\{
\mathbf{H}_{\mathrm{CF}}
\right\},
\label{eq:global_rankone_projection}
\end{equation}
where $\mathcal{P}_{1}\{\cdot\}$ denotes the best rank-one
approximation in the Frobenius-norm sense.
\textcolor{black}{Specifically, if
$(\sigma_{\mathrm{CF},1},\mathbf{u}_{\mathrm{CF},1},
\mathbf{v}_{\mathrm{CF},1})$ is a dominant singular triplet of
$\mathbf H_{\mathrm{CF}}$, then}
\begin{equation}
\color{black}
\mathbf{H}_{\mathrm{r1}}
=\underset{\substack{\mathbf X\in\mathbb{C}^{T\times N}\\
\operatorname{rank}(\mathbf X)\leq 1}}{\arg\min}
\left\|\mathbf H_{\mathrm{CF}}-\mathbf X\right\|_{F}^{2}
=\sigma_{\mathrm{CF},1}
\mathbf{u}_{\mathrm{CF},1}\mathbf{v}_{\mathrm{CF},1}^{H}.
\label{eq:rankone_projection_objective}
\end{equation}
\textcolor{black}{This projection is the nearest rank-one matrix to the
assembled factor. The subsequent update is a single conditional refinement.}
\textcolor{black}{For fixed $\mathbf H_{\mathrm{r1}}$, define the
mode-3 design matrix}
\begin{equation}
\color{black}
\mathbf A_{\mathrm{r1}}
\triangleq
\left(\mathbf H_{\mathrm{r1}}\diamond\mathbf I_N\right)^T
\in\mathbb{C}^{N\times NT}.
\label{eq:Ar1_definition}
\end{equation}
\textcolor{black}{The columns of
$\mathbf H_{\mathrm{r1}}\diamond\mathbf I_N$ have disjoint support,
which gives}
\begin{equation}
\color{black}
\mathbf A_{\mathrm{r1}}\mathbf A_{\mathrm{r1}}^{H}
=\operatorname{diag}\left\{
\|\mathbf h_{\mathrm{r1},1}\|_{2}^{2},\ldots,
\|\mathbf h_{\mathrm{r1},N}\|_{2}^{2}
\right\}.
\label{eq:Ar1_gram}
\end{equation}
\textcolor{black}{Here, $\mathbf h_{\mathrm{r1},n}$ denotes the $n$th
column of $\mathbf H_{\mathrm{r1}}$. Thus, if every such column is
nonzero, $\mathbf A_{\mathrm{r1}}$ has full row rank and the
unconstrained LS estimate of the morphing factor is unique:}
\begin{equation}
\color{black}
\widetilde{\mathbf{E}}_{\mathrm{CF}}
=\underset{\mathbf E\in\mathbb C^{M\times N}}{\arg\min}\;
\left\|[\mathcal Y]_{(3)}-\mathbf E\mathbf A_{\mathrm{r1}}\right\|_F^2
=
[\mathcal{Y}]_{(3)}
\mathbf A_{\mathrm{r1}}^{\dagger}.
\label{eq:E_refinement}
\end{equation}
\textcolor{black}{This is ordinary, unweighted conditional least squares
(and is conditionally maximum likelihood under i.i.d. white Gaussian
noise). 
}
Define $\mathbf{D}_{\mathrm{CF}}
\triangleq
\operatorname{diag}
\{(\widetilde{\mathbf{E}}_{\mathrm{CF}}(1,:))^{T}\}$,
the final closed-form factor estimates are then
\begin{equation}
\widehat{\mathbf{E}}
=
\widetilde{\mathbf{E}}_{\mathrm{CF}}
\mathbf{D}_{\mathrm{CF}}^{-1},
\qquad
\widehat{\mathbf{H}}
=
\mathbf{H}_{\mathrm{r1}}
\mathbf{D}_{\mathrm{CF}}.
\label{eq:final_factors}
\end{equation}
\textcolor{black}{Provided every diagonal entry of
$\mathbf D_{\mathrm{CF}}$ is nonzero, this diagonal rescaling enforces
the reference-state constraint without changing the fitted tensor:
}
\begin{equation}
\color{black}
\widehat{\mathbf E}
\left(\widehat{\mathbf H}\diamond\mathbf I_N\right)^T
=\widetilde{\mathbf E}_{\mathrm{CF}}\mathbf A_{\mathrm{r1}}.
\label{eq:normalization_invariance}
\end{equation}
\textcolor{black}{In noisy data, small diagonal entries likewise make
this normalization ill-conditioned and should be treated as a
degenerate estimate.}
Consequently,
\begin{equation}
\operatorname{rank}(\widehat{\mathbf{H}})=1,
\qquad
\widehat{\mathbf{E}}(1,:)=\mathbf{1}_{N}^{T}.
\end{equation}
\textcolor{black}{The algebraic factor recovery stage is deterministic and noniterative. Its main operations are $N$ independent dominant singular vector computations, one global rank-one projection, and one structured least-squares update; physical angle extraction is subsequently performed by the manifold-matching step.}

\textit{Remark 1 (Recoverability and complexity):}
The proposed estimator does not rely on generic Kruskal-rank
uniqueness. Since the known identity factor $\mathbf I_N$ decouples
the mode-1 unfolding into $N$ rank-one factorizations
$\mathbf Q_n=\mathbf h_n\boldsymbol{\epsilon}_n^T$, their scalar
ambiguities are removed by the reference condition
$[\boldsymbol{\epsilon}_n]_1=1$. Recovery of
$\boldsymbol{\alpha}$ further requires
$\operatorname{rank}(\mathbf S)=K$, i.e., $T\ge K$.
\textcolor{black}{For algebraic factor recovery, the main cost is computing the leading singular triplet of each of the $N$ matrices of size $T\times M$, with complexity $\mathcal O(NTM\min\{T,M\})$. This is followed by one global rank-one projection and one structured least-squares refinement; the stated complexity excludes the subsequent angular-manifold search.}

\begin{algorithm}[t]
	\caption{Closed-Form FIM-PARAFAC Estimator}
	\label{alg:cf_parafac}
	\begin{algorithmic}[1]
		\Require $\mathcal{Y}$, $\mathbf{S}$
		\Ensure $\widehat{\mathbf{H}}$, $\widehat{\mathbf{E}}$,
		$\widehat{\boldsymbol{\alpha}}$,
		$(\widehat{\theta},\widehat{\phi})$
		
		\State Form $\mathbf{Q}=[\mathcal{Y}]_{(1)}^{T}$.
		\For{$n=1,\ldots,N$}
		\State Reshape $\mathbf{q}_n$ into $\mathbf{Q}_n$
		and compute its dominant singular triplet.
		\State Obtain
		$\widetilde{\mathbf h}_n$ and
		$\widetilde{\boldsymbol{\epsilon}}_n$ from \eqref{eq:svd_factor}.
		\State Normalize the pair using \eqref{eq:column_normalization}.
		\EndFor
		\State Form $\mathbf{H}_{\rm CF}$ and apply
		$\mathbf{H}_{\rm r1}=\mathcal{P}_1\{\mathbf{H}_{\rm CF}\}$.
		\State Refine and normalize $\mathbf{E}$ using
		\eqref{eq:E_refinement}--\eqref{eq:final_factors}.
		\State Recover $\widehat{\boldsymbol{\alpha}}$ using
		\eqref{eq:gain_recovery}.
		\State Estimate $(\widehat{\theta},\widehat{\phi})$
		using \eqref{eq:angle_matching}.
	\end{algorithmic}
\end{algorithm}

\subsection{Physical Parameter and Channel Recovery}

After estimating the algebraic factors
$\widehat{\mathbf{H}}$ and $\widehat{\mathbf{E}}$, the physical
channel parameters can be recovered by exploiting the rank-one
structure
$\mathbf{H}=\mathbf{z}\mathbf{a}^{T}$. Let $\widehat{\mathbf{H}}=\mathbf{U}_{H}\boldsymbol{\Sigma}_{H}\mathbf{V}_{H}^{H}$,
and denote its dominant singular triplet by
$(\sigma_{H,1},\mathbf{u}_{H,1},\mathbf{v}_{H,1})$.
A compatible pair of estimates is
\begin{equation}
\widetilde{\mathbf{z}}
=
\sqrt{\sigma_{H,1}}\mathbf{u}_{H,1},
\qquad
\widetilde{\mathbf{a}}
=
\sqrt{\sigma_{H,1}}\mathbf{v}_{H,1}^{*}.
\end{equation}
\textcolor{black}{A residual scalar ambiguity remains between $\mathbf{z}$ and $\mathbf{a}$. Since the first element of the reference UPA steering vector is $[\mathbf{a}]_{1}=1$, define $c\triangleq[\widetilde{\mathbf{a}}]_{1}$. The reference-normalized estimates are $\widehat{\mathbf{a}}=\widetilde{\mathbf{a}}/c$ and $\widehat{\mathbf{z}}=c\widetilde{\mathbf{z}}$.}
\textcolor{black}{The individual user gains are then recovered from the known pilot matrix as}
\begin{equation}
\widehat{\boldsymbol{\alpha}}
=
(\mathbf{S}^{T})^{\dagger}
\widehat{\mathbf{z}}.
\label{eq:gain_recovery}
\end{equation}
Unique recovery of the user gains requires $\operatorname{rank}(\mathbf{S})=K$, and $T\geq K$.
Finally, the common angular parameters are extracted by jointly
matching the recovered spatial and morphing factors to the
calibrated FIM manifold. Defining
\begin{equation}
\mathbf{E}(\theta,\phi)
=
[\mathbf{e}_{1}(\theta,\phi),
\ldots,
\mathbf{e}_{M}(\theta,\phi)]^{T},
\end{equation}
the angular estimate is obtained from
\begin{align}
(\widehat{\theta},\widehat{\phi})
=
\arg\min_{(\theta,\phi)\in\Omega}
\;&
1-
\frac{
	\left|
	\mathbf{a}_{\mathrm{UPA}}^{H}(\theta,\phi)
	\widehat{\mathbf{a}}
	\right|^{2}
}{
	\|\mathbf{a}_{\mathrm{UPA}}(\theta,\phi)\|_{2}^{2}
	\|\widehat{\mathbf{a}}\|_{2}^{2}
}
\nonumber\\
&+
\lambda_{E}
\frac{
	\|
	\widehat{\mathbf{E}}
	-
	\mathbf{E}(\theta,\phi)
	\|_{F}^{2}
}{
	\|\widehat{\mathbf{E}}\|_{F}^{2}
},
\label{eq:angle_matching}
\end{align}
where $\Omega$ denotes the admissible angular region and
$\lambda_E\geq0$ controls the contribution of the morphing
factor. The channel of user $k$ under the $m$th morphing configuration
is finally reconstructed as
\begin{equation}
\widehat{\mathbf{h}}_{k,m}
=
\widehat{\alpha}_{k}
\mathbf{a}_{m}
(\widehat{\theta},\widehat{\phi}),
\,\,
k=1,\ldots,K,\,\,
m=1,\ldots,M.
\label{eq:channel_reconstruction}
\end{equation}
 
\begin{figure*}[!t]
	\centering
	\begin{minipage}[t]{0.48\textwidth}
		\centering
		\includegraphics[height=0.20\textheight,
		keepaspectratio]{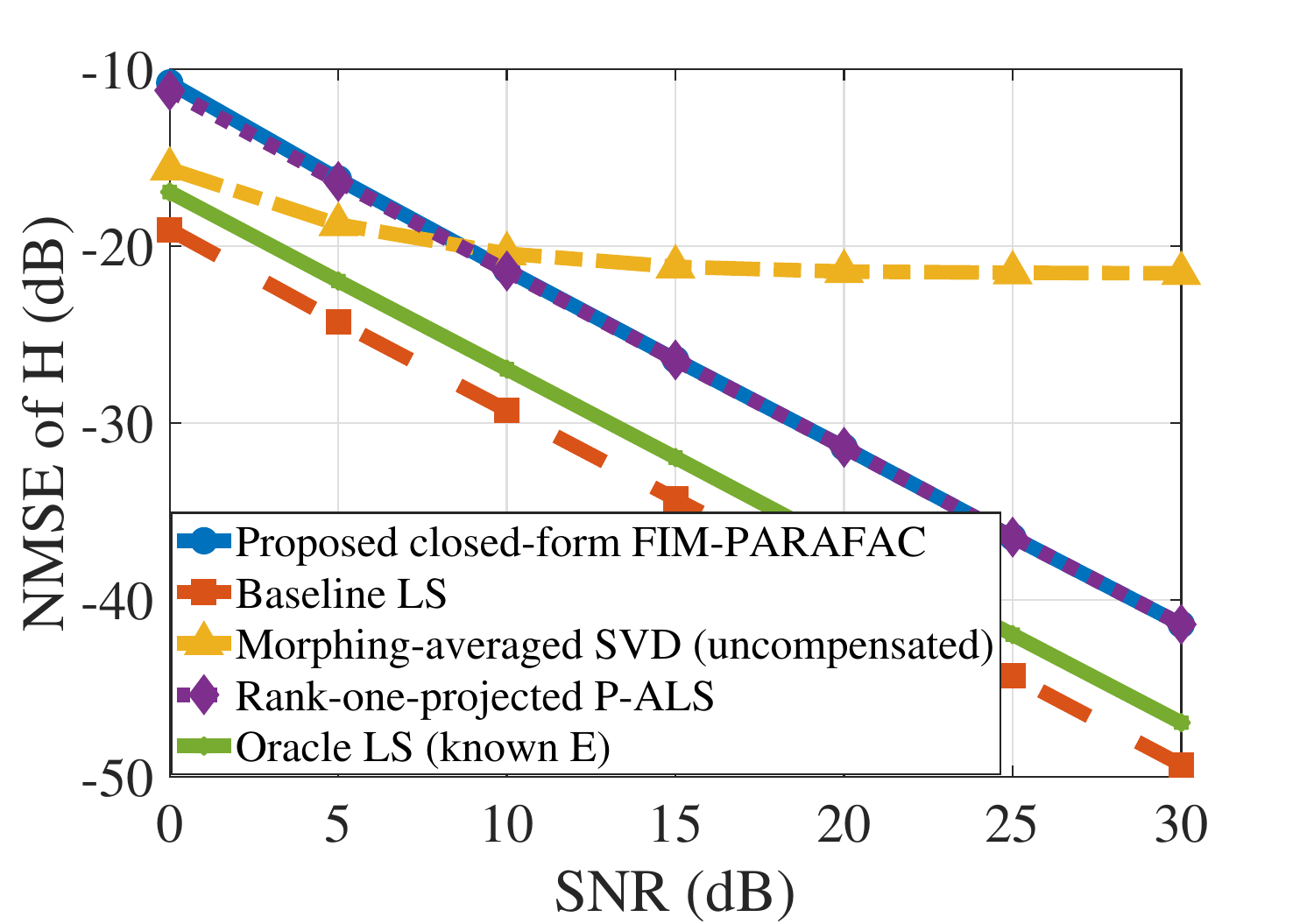}
		\vspace{-1mm}
		\centerline{(a)}
	\end{minipage}
	\hfill
	\begin{minipage}[t]{0.48\textwidth}
		\centering
		\includegraphics[height=0.20\textheight,
		keepaspectratio]{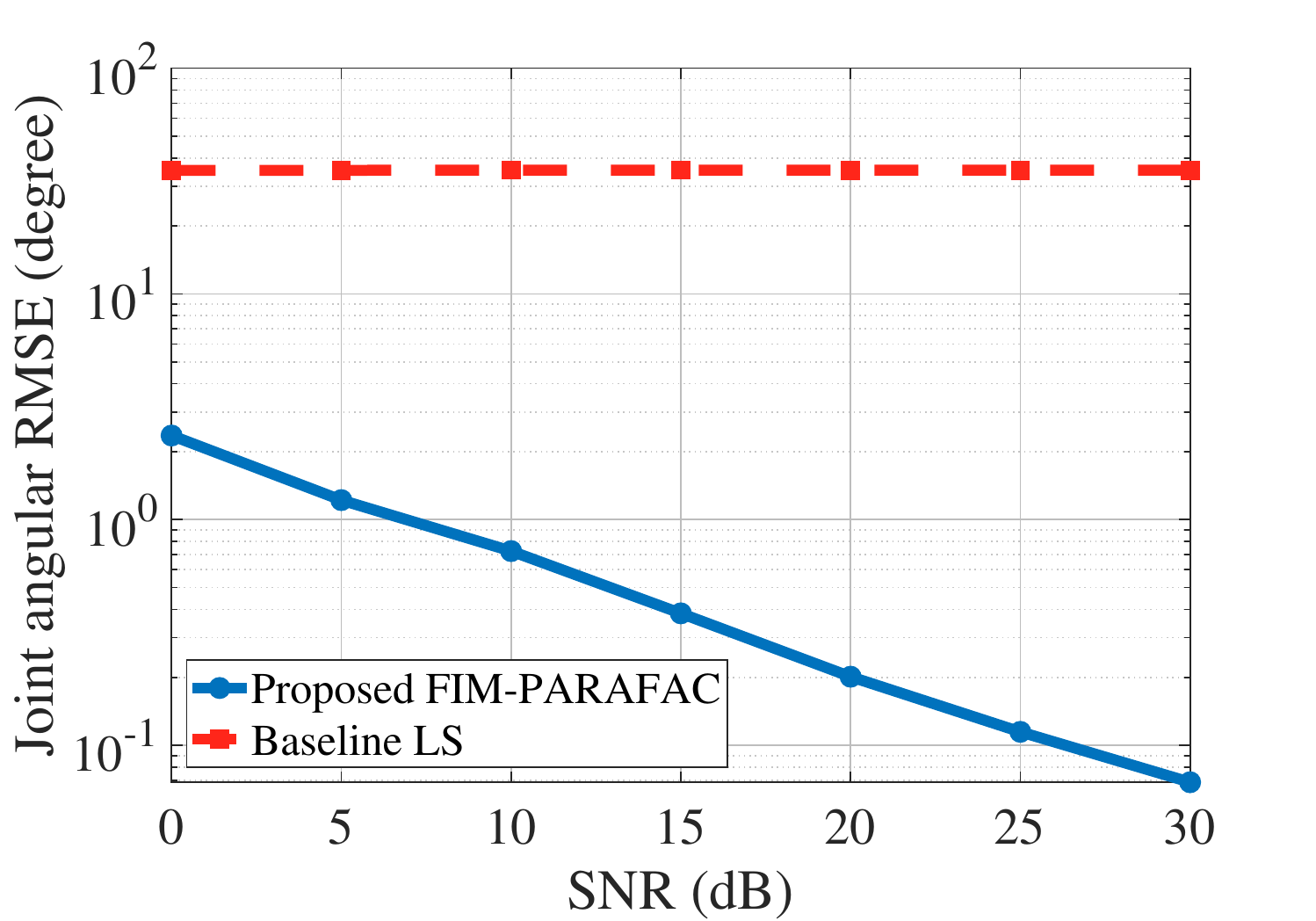}
		\vspace{-1mm}
		\centerline{(b)}
	\end{minipage}
	\caption{Performance versus SNR for $N=16$, $K=3$,
		$T=16$, and $M=6$: (a) NMSE of $\mathbf H$;
		(b) joint angular RMSE.}
	\label{fig:snr_results}
	\vspace{-2mm}
\end{figure*}

\section{Simulation Results}
\label{sec:results}

We consider an FIM with $N=N_xN_z=16$ elements, where
$N_x=N_z=4$ and $d_x=d_z=\lambda/2$ with $\lambda=1$.
Unless otherwise stated, $K=3$, $T=16$, and $M=6$, and all
users share the common angular pair
$(\theta,\phi)=(35^\circ,25^\circ)$.
The channel gains follow
$\alpha_k\sim\mathcal{CN}(0,1/K)$, while the nonreference
element displacements are independently drawn from
$\mathcal{U}[-\lambda/4,\lambda/4]$.
Deterministic DFT pilots satisfying
$\mathbf{S}\mathbf{S}^{H}=T\mathbf{I}_{K}$ are employed, and
all results are averaged over $N_{\rm MC}=2000$ realizations.
For each realization, the noise variance is set as
$\sigma_w^2=P_{\mathcal X}10^{-\mathrm{SNR}/10}$, where
$P_{\mathcal X}=\|\mathcal X\|_F^2/(NMT)$.
The NMSE is averaged over all Monte Carlo realizations and
reported in dB. The joint angular RMSE is
\begin{equation}
\mathrm{RMSE}_{\rm ang}
=
\sqrt{
	\frac{1}{2N_{\rm MC}}
	\sum_{i=1}^{N_{\rm MC}}
	\left[
	(\Delta\theta^{(i)})^2+
	(\Delta\phi^{(i)})^2
	\right]},
\label{eq:angular_rmse}
\end{equation}
where $\Delta\phi^{(i)}$ is wrapped to
$[-180^\circ,180^\circ)$.
For a fair comparison, the equal-overhead rigid-panel LS
baseline uses $T_{\rm rigid}=MT$ pilots under the undeformed
configuration, with
$\mathbf{S}_{\rm rigid}=[\mathbf{S},\ldots,\mathbf{S}]$.

\subsection{Estimation and Angular-Recovery Performance}

\textcolor{black}{Fig.~\ref{fig:snr_results}(a) compares the normalized mean squared error (NMSE) of the equivalent rank-one matrix $\mathbf H$. The proposed structured estimator improves as SNR increases, whereas the morphing-averaged singular value decomposition (SVD) exhibits an error floor because it does not account for configuration-dependent phase responses. The oracle LS curve serves as a lower bound. At equal pilot-symbol overhead, the rigid-panel LS attains lower $\mathbf H$-NMSE because all $MT$ pilots observe the same fixed spatial response and thus provide greater noise averaging. The proposed method instead distributes those pilots across multiple morphing configurations and estimates the morphing-dependent factor jointly. This tradeoff highlights the distinct benefit of morphing thanks to improved angular identifiability.}

\textcolor{black}{Fig.~\ref{fig:snr_results}(b) demonstrates the value of morphing diversity for physical-angle recovery. The proposed estimator's angular root mean square error (RMSE) decreases with SNR, whereas the rigid-panel LS retains an error floor of approximately $35^\circ$. For the undeformed $xz$-plane array, $\mathbf a_{\rm UPA}(\theta,\phi)=\mathbf a_{\rm UPA}(\theta,-\phi)$, so the sign of $\phi$ is unidentifiable. With $\phi=25^\circ$, the indistinguishable solution $-25^\circ$ incurs a $50^\circ$ azimuth error and, under the joint RMSE definition in \eqref{eq:angular_rmse}, yields approximately $35.36^\circ$. Out-of-plane morphing adds a phase dependence on $\sin\theta\sin\phi$, breaking this ambiguity through joint spatial--morphing manifold matching.}

\subsection{Impact of Morphing Diversity and Model Mismatch}

\begin{figure*}[!t]
	\centering
	\begin{minipage}[t]{0.48\textwidth}
		\centering
		\includegraphics[height=0.21\textheight,
		keepaspectratio]{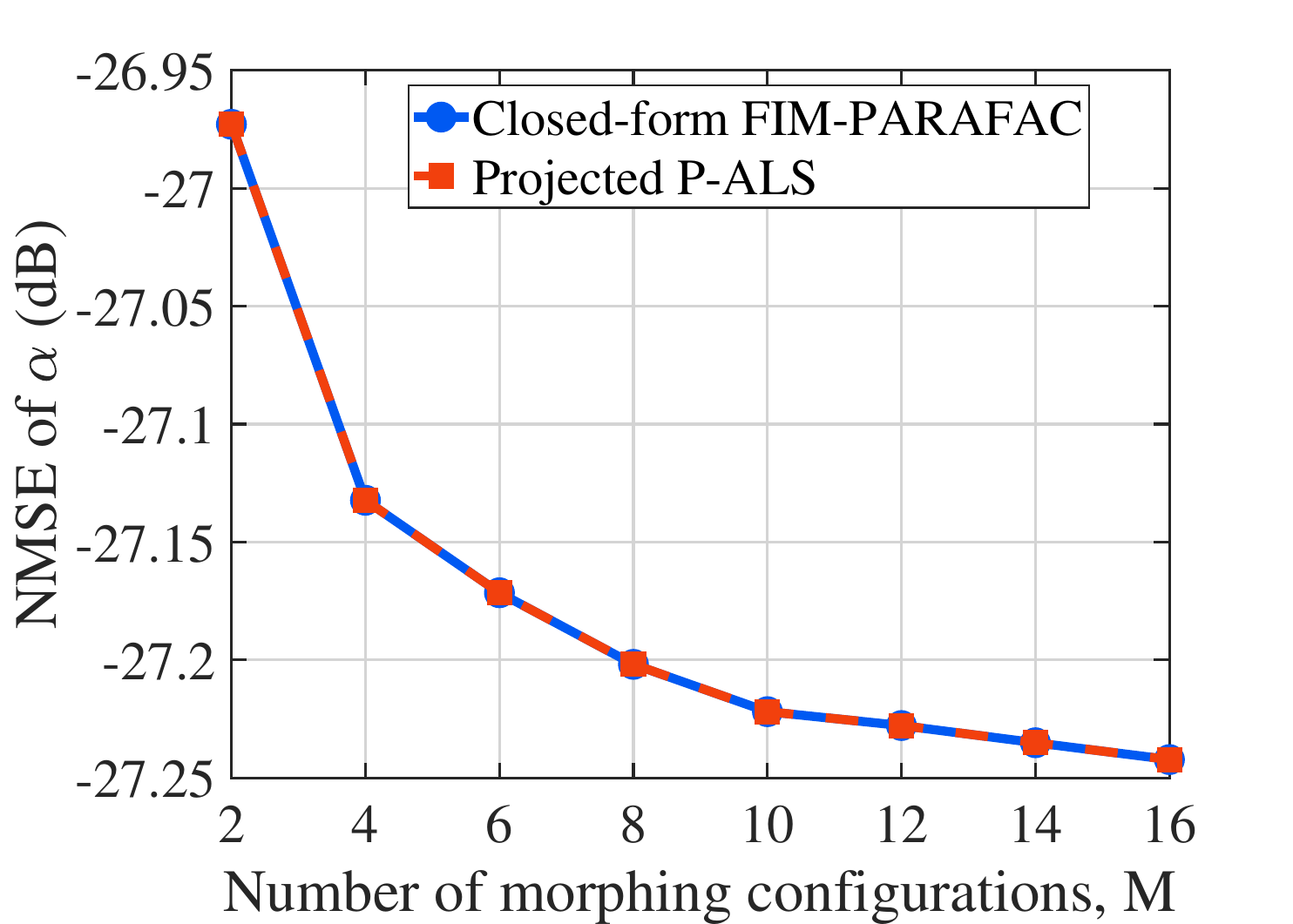}
		\vspace{-1mm}
		\centerline{(a)}
	\end{minipage}
	\hfill
	\begin{minipage}[t]{0.48\textwidth}
		\centering
		\includegraphics[height=0.21\textheight,
		keepaspectratio]{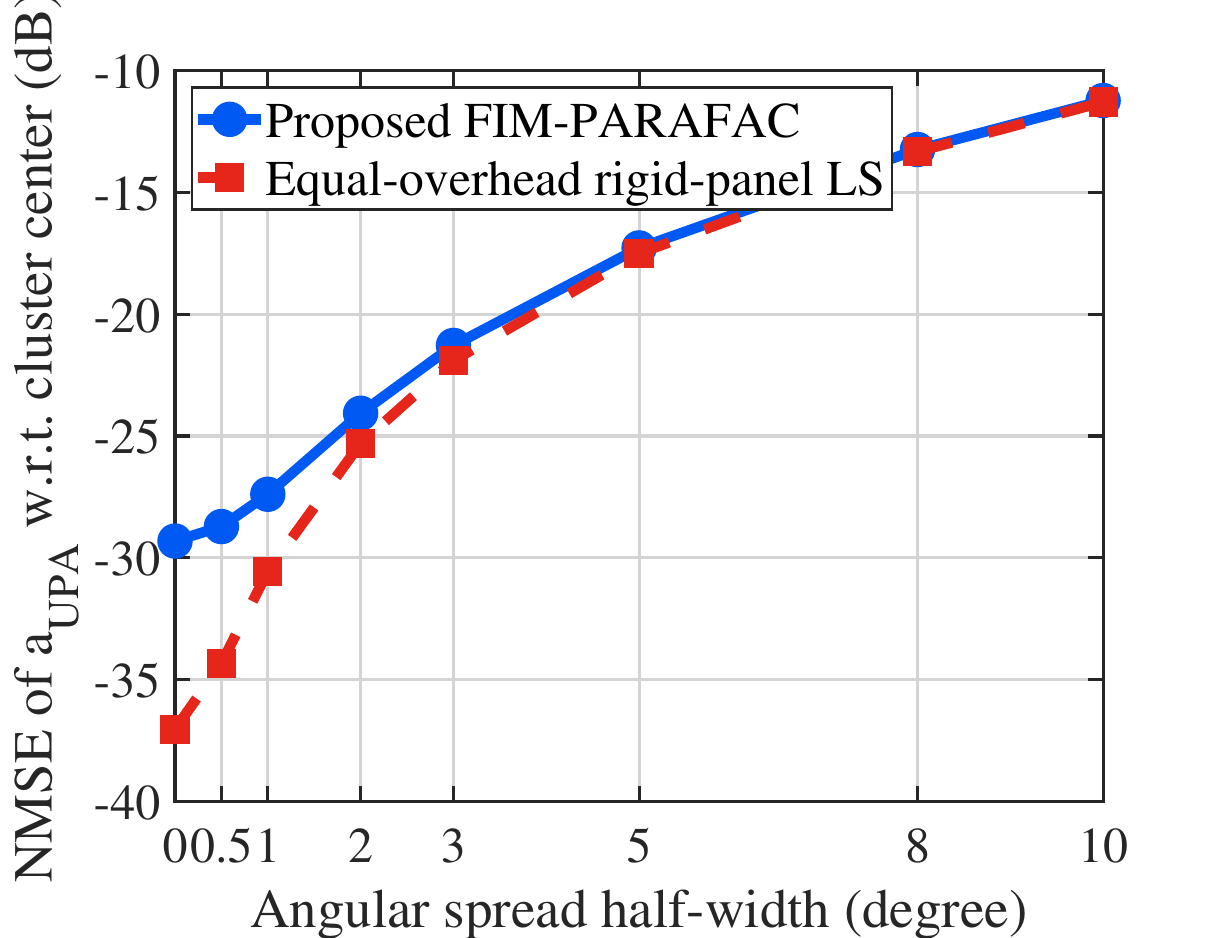}
		\vspace{-1mm}
		\centerline{(b)}
	\end{minipage}
	\caption{Impact of morphing diversity and angular spread:
		(a) NMSE of $\boldsymbol{\alpha}$ versus $M$ at
		$\mathrm{SNR}=15$ dB;
		(b) noise-free common-angular-response model mismatch
		versus angular-spread half-width.}
	\label{fig:morphing_results}
	\vspace{-2mm}
\end{figure*}

\textcolor{black}{Fig.~\ref{fig:morphing_results}(a) shows that the NMSE of $\boldsymbol{\alpha}$ decreases as $M$ grows, since additional calibrated configurations provide more observations for structured factor recovery. The gains diminish at larger $M$. The closed-form and projected alternating estimators perform nearly identically under the matched common angular-response model, indicating that the noniterative factor-recovery stage captures most of the available benefit. Because each configuration uses $T$ pilot symbols, increasing $M$ also increases the total training overhead in this experiment.}

\textcolor{black}{Fig.~\ref{fig:morphing_results}(b) assesses the common angular-response model under nonzero angular spread. It plots the normalized, noise-free residual between the true multiuser observation tensor and its best structured common-angle PARAFAC approximation. At zero spread, all users share the same spatial and morphing responses, so the residual approaches numerical precision. As the spread grows, user-dependent responses cannot be represented by the single rank-one factor $\mathbf{H}=\mathbf{z}\mathbf{a}^{T}$, and the irreducible model mismatch increases. This experiment therefore indicates the regime in which the proposed formulation is appropriate: users with a shared dominant direction or a sufficiently compact angular cluster.}

\section{Conclusion}

\textcolor{black}{This letter introduced a structure-aware PARAFAC estimator for multiuser uplink channel estimation across calibrated FIM morphing configurations under a common-angular-response model. The estimator exploits the known identity factor, the global rank-one pilot-weighted channel structure, and the undeformed reference state to recover the algebraic factors deterministically without alternating updates; calibrated-manifold matching then yields the common angles, user gains, and configuration-dependent channels. Simulations demonstrate accurate factor recovery, show that out-of-plane morphing resolves the azimuth-sign ambiguity of an undeformed planar aperture, and quantify the model mismatch that arises with nonzero angular spread.}

\renewcommand{\baselinestretch}{.88}


\end{document}